\documentclass[final,5p,times,twocolumn,number]{elsarticle}

\usepackage{graphics}

\usepackage{amssymb}
\usepackage{amsthm}

\usepackage{dcolumn}
\usepackage{threeparttable}
\usepackage{bm}

\journal{Chemical Physics Letters}

\begin{document}

\begin{frontmatter}



\title{A consistent description of the iron dimer spectrum with a 
correlated single-determinant wave function}


\author[poly]{Michele Casula}
\ead{michele.casula@gmail.com}

\author[sissa,democritos]{Mariapia Marchi}

\author[sissa]{Sam Azadi}

\author[sissa,democritos]{Sandro Sorella}

\address[poly]{Centre de Physique Th\'eorique, Ecole Polytechnique, CNRS, 91128 Palaiseau, France}
\address[sissa]{SISSA, International School for Advanced Studies, 34014, Trieste, Italy}
\address[democritos]{DEMOCRITOS National Simulation Center, 34014, Trieste, Italy}

\begin{abstract}
We study the iron dimer by using an accurate ansatz for quantum chemical 
calculations based on a simple variational wave function,  
defined by a single geminal expanded in molecular orbitals
and combined with a real space correlation factor.
By means of this approach
we predict that, contrary to previous expectations, the neutral ground state is 
$^7 \Delta$ while the ground state of the anion is $^8 \Sigma_g^-$, hence explaining in a simple way 
a long standing controversy in the interpretation of the experiments.
Moreover, we characterize consistently the states 
seen in the photoemission spectroscopy by Leopold \emph{et al.}.\cite{leopold86} 
It is shown that the non-dynamical correlations included in the geminal expansion are relevant to 
correctly reproduce the energy ordering of the low-lying spin states.
\end{abstract}





\end{frontmatter}


\section{Introduction}
\label{introduction}
The iron dimer is a puzzling molecule. Indeed, explaining all the 
related experimental findings in a consistent theoretical frame is very hard. 
Moreover, this dimer represents 
a testcase to check the ability of a method to capture the physics of
the transition metal compounds, since it includes all their difficulties,
namely a strong electron correlation in the nearly half-filled $d$ orbitals, and
a non trivial ordering of the low-lying energy states which differ
by their spin.

In 1986 Leopold \emph{et al.} carried out an experiment\cite{leopold86} of
negative ion photoelectron spectroscopy (PES), with the aim of
studying the low-lying electronic states of Fe$_2$.
A sample of Fe$^-_2$ is prepared and excited by an incoming
photon. The spectrum of Fe$_2$ appears remarkably simple, with only two peaks, 
corresponding to the excitations from the Fe$^-_2$ ground state to those Fe$_2$ states allowed by the selection rules,
which implies that the total angular momentum of the final state cannot change by more than one.
Both states reveal the same vibrational frequency and bond length. Few years
later, Leopold\cite{leopold88} 
argued that the simplest explanation of these data is to
admit that the $^9\Sigma_g^-$ is the ground state of Fe$_2$, 
and the $^8\Sigma_u^-$ is the ground state of Fe$^-_2$.
This interpretation is based on the hypothesis
that the two-band system observed in the Fe$^-_2$ spectrum 
is due to the detachment from a $4s$-like
molecular orbital (MO).
Therefore, if one supposes that the ground state of Fe$^-_2$ is
$^8\Sigma_u^-$, its configuration turns out to be
$\sigma^2_g(4s)~\sigma^{*,2}_u(4s) 3d^{13}$, and 
the $4s$ electron detachment would
produce two possible states, with the same orbital configuration
$\sigma^2_g(4s)~\sigma^{*,1}_u(4s) 3d^{13}$ but with the $\sigma^*_u(4s)$
of high ($^9\Sigma_g^-$) or low ($^7\Sigma_g^-$) spin 
coupled to the remaining $3d^{13}$ electrons. These two states would
correspond to the first and second peak of the spectrum,
respectively, and display the same structural properties, 
the only difference being the spin coupling.
However, this interpretation disagrees with electron spin resonance experiments,\cite{baumann84}
which fail to observe the Fe$_2$ and therefore supports
the idea of an orbitally degenerate ground state, like the
$^7\Delta_u$ configuration, unless the iron dimer features a 
large zero-field split (larger than 8 $cm^{-1}$), thus producing an energy split 
not detectable by the experimental setup. 
However, such a large magnetic split is unusual for that kind of molecules.

From the theoretical side, 
some numerical studies based mainly on density functional theory (DFT) methods with
various functionals\cite{castro93,castro94,chretien02,gutsev03} 
and multi reference configuration interaction (MRCI) 
calculations\cite{tomonari88} yielded the $^7\Delta_u$
as the ground state of Fe$_2$, while more recent 
MRCI calculations\cite{hubner02} and DFT studies with coupled
cluster\cite{irigoras03} and $+U$\cite{kulig07} corrections supported
the idea that its ground state is $^9\Sigma_g^-$.
Those methods gave a $^8\Delta_g$  ($^8 \Sigma_u^-$) ground state for the anion
whenever a  $^7\Delta_u$ ($^9\Sigma_g^-$) ground state was found for the neutral dimer.

In this Letter, we tackle the study of the iron dimer 
by means of quantum Monte Carlo (QMC) simulations based on 
the resonating valence bond (RVB) wave function,\cite{pauling,pwanderson}
which is a correlated single-determinant ansatz 
successfully used in previous calculations.\cite{casula1,casulamol,pfaff1,pfaff2}
Here, we use an extension of the RVB picture
that is  based on a MO expansion of the singlet state in the 
determinant.
By setting the number of MO's to a value such that 
a Jastrow correlated single determinant (SD) 
wave function is recovered
for each fragment in the atomization limit, 
we obtain a description of the bond which is remarkably accurate.
This ansatz has been tested on a set of first-row atom dimers,
such as F$_2$, N$_2$, Be$_2$, and C$_2$,
where our calculations yielded results 
for the binding energy and the equilibrium distance
very close to the experimental values,\cite{to_be_published}
much better than those obtained in previous calculations
based on standard Jastrow-SD (JSD) wave functions.\cite{grossman}

\section{Method}
\label{method}

We consider a RVB wave function for $N$ electrons,
where for simplicity we take $N_\uparrow\ge N_\downarrow$, 
and $N_\uparrow$ ($N_\downarrow$) is the number of spin up (down) electrons.
The wavefunction is given by the product of a determinantal part and a Jastrow
correlation factor.
The determinantal part is the antisymmetrized product 
of singlet pairs. Each pair is described by a symmetric 
$\phi(\vec r,\vec r^\prime) =\phi(\vec r^\prime,\vec r)$ 
orbital function. 
In order to define a pure spin state with total spin 
$S=|N_\uparrow-N_\downarrow|/2$ and maximum spin projection $S^{tot}_z=S$,
we take $N_\downarrow$ singlet pairs 
and $2S$ unpaired orbitals $\phi_j(\vec r)$ for spin up electrons, 
and apply the antisymmetrization operator ${\cal A}$ to obtain a 
consistent fermionic wave function: 
\begin{equation} \label{AGP}
\Phi_N(\vec R)= {\cal A }  \prod\limits_{i=1}^{N_\downarrow} 
\phi (\vec r^\uparrow_i, \vec r^\downarrow_i ) \prod_{j=N_\downarrow+1}^{N_\uparrow} 
\phi_j (\vec r^\uparrow_j ),
\end{equation}
where $\Phi_N$ is the $N$-electron wave function and $\vec R$ indicates 
the corresponding $3N$-dimensional vector of coordinates,
$\vec R= \left\{ \vec r_1^\uparrow, \cdots, \vec r_{N_\uparrow}^\uparrow, r_1^\downarrow, \cdots ,  \vec r_{N_\downarrow}^\downarrow \right\} $.
The wave function in Eq.~(\ref{AGP}) is also called antisymmetrized geminal product (AGP) 
and can be computed by means of a single determinant (see Ref.~\cite{casula1} and references therein).

As said, the total spin of $\Phi_N(\vec R)$ is definite.
We also impose 
all possible symmetries to be satisfied, including  
angular momentum and spatial reflections.

The Jastrow correlation factor is the other important ingredient of the wavefunction.
Its generally adopted form reads:
\begin{equation}\label{jastrow}
J(x) = \exp \left(  \sum \limits_{i<j} f ( \vec r_i ,\vec r_j)  \right), 
\end{equation}
where $f (\vec r ,\vec r^\prime)$ 
is a function of two electron coordinates.\cite{casulamol}
The Jastrow term accounts for the electron-electron repulsion
and suppresses configurations
with overlapping valence bonds, which would
lead to a too large electron density around an atom, with an 
increase in the total energy. 

As any function of two coordinates,  the pairing function $\phi$ as 
well as the correlation function $f$ in the Jastrow term can be 
expanded in terms of single particle orbitals. In particular, 
the pairing function reads:
\begin{equation} \label{pairing}
\phi( \vec r, \vec r^\prime) = \sum\limits_{j=1}^{n-2S} 
\lambda_j \phi_j( \vec r) \phi_j( \vec r^\prime), 
\end{equation}
where $n$ is large enough, and $\{ \phi_j \}$ is an orthogonal 
single particle basis set\footnote{The orbitals $\phi_i$ can be MO's 
expanded  in terms of atomic orbitals $\varphi_{a,j}$ where $a$ indicates the atomic center 
and $j$ the type: 
$\phi_i (\vec r) = \sum_{a,j} \chi^i_{a,j} \varphi_{a,j} (\vec r)$.
The coefficients $\chi^i_{a,j}$, as well as the weights $\lambda_j$, 
can be used as variational parameters defining the geminal in 
Eq.~(\ref{pairing})}, which reaches 
its complete basis set limit for $n\to \infty$.
Notice also that in these notations we assume that the $2S$ unpaired orbitals 
$\phi_j$ correspond to the indexes: $ n-2S+1 \le j \le n$.

The single particle orbitals $\phi_j$ can be conveniently chosen as
the MO's obtained with a conventional restricted Hartree-Fock (RHF)
calculation. Indeed the MO basis 
allows us to write Eq.~(\ref{pairing})
in a diagonal form equivalent 
to a more involved matrix form when the 
MO's are  developed in an atomic basis set.\cite{casulamol}
By truncating the expansion in Eq.~(\ref{pairing}) 
to a number of MO's $n$ equal to the number 
of electron pairs and unpaired orbitals, namely $n=N_\uparrow$,
one recovers the usual RHF theory,  
because the antisymmetrization operator ${\cal A}$  clearly singles out only 
one Slater determinant. Moreover, the MO weights  $\lambda_j$ 
affect only an overall prefactor of this Slater determinant, 
so that their actual values are irrelevant in this case. 
However, the pairing function is generally 
not limited to have only $N_\downarrow$ non vanishing 
eigenvalues $\lambda_j$.
Therefore, the RVB wave function represents a clear extension of the RHF theory,
not only for the  presence of the Jastrow factor, which considerably improves 
the dynamical correlations, but mainly  
because its determinantal part goes beyond RHF 
when $n>N_\uparrow$, by including also non-dynamical correlations.
Quite generally, a gain in energy and a more accurate 
calculation are expected whenever $n > N_{\uparrow}$. 

In this Letter, we use in all calculations a number $n^*$
of MO's that is enough to have 
a fully symmetric state that connects the compound at rest to the 
atoms at large distance, where a fragmented 
JSD wave function is recovered.
A larger value of $n$ certainly leads to a lower value of the total energy,
but may improve much more the atomic energies, rather than the bonding.

Clearly, whenever $n=n^*$ the atomization energy has to be 
referenced to the JSD calculation, even when better energies are provided by 
the RVB for the atoms.\cite{casula1}
It is important to remark here 
that, upon  stretching the molecule to the 
atomization limit, some symmetries of the atomic wave function 
are not recovered.
For instance, for a diatomic molecule such as iron, 
 the total angular momentum is conserved only for the rotations 
around the molecular axis, so that the total angular momentum of the 
atomic fragments could not be definite within this ansatz.
Therefore, by assuming that this error does not affect the 
molecular bond, 
a correction to the atomic reference is necessary, given by 
the energy difference between 
the hybrid SD state reached upon streaching the RVB wave function
and the fully symmetric JSD atomic state computed
with the same primitive basis as the one used for the molecule. 

Our RVB wave function is the input for QMC simulations. 
We start from its optimization, and 
then perform variational Monte Carlo (VMC) and diffusion Monte Carlo 
(DMC) simulations,\cite{mitas} the latter within its recent lattice regularized implementation
(LRDMC).\cite{lrdmc} The optimization method used here is based on the
calculations of the Hamiltonian matrix elements in the space spanned by the wave function\cite{cyrus,rocca},
recently developed to perform a constrained energy minimization with fixed $n$.\cite{to_be_published}

\section{Results}
We compute the properties of the  
$^7\Delta_u$, $^7\Sigma_g^-$, $^9\Sigma_g^-$ states for the neutral dimer,
with the method described in Sec.~\ref{method}, namely with the idea that 
only a well controlled dissociation limit can lead to reliable predictions for the energetics 
in the bonding region. 
We also compute the $^8\Sigma_u^-$ and $^8\Delta_g$ states of the anion 
to make a direct comparison with the PES.

For all our calculations, we employ a neon-core pseudopotential, 
in order to avoid the chemically inert core
electrons of the iron atom, and speed up the QMC simulations. We choose the
Dolg's pseudopotential,\cite{dolg87} which has previously proven to be
reliable at least for atomic QMC calculations.\cite{mitas94} 
We use a $[8s5p6d/2s1p1d]$ contracted Gaussian basis set, which
leads to a space spanned by 4 $\sigma$, 4 $\sigma^*$, 4 $\pi$, 4 $\pi^*$, 2 $\delta$, and 2 $\delta^*$ MO's,
where we need to accommodate 32 electrons for the neutral dimer.
Our primitive basis set is quite compact. However, we double checked the
convergence in the energy differences for atomic calculations by extending the basis set 
up to $8s5p6d3f$.
A much smaller basis was used for the Jastrow factor, 
because this allows for a more efficient energy optimization.
On the other hand, 
the essentially exact contribution of Jastrow-type dynamical correlations,
which do not change the phases of the wave function, can be very accurately 
obtained with DMC or LRDMC. 
The (LR)DMC approach can be seen as a stochastic optimization of a much more general Jastrow factor
which keeps fixed the nodes of the RVB wave function.

For the  $^9\Sigma_g^-$, $^8\Sigma_u^-$, and $^8\Delta_g$ states  $n^*=N_\uparrow$ 
is such that the AGP is an HF Slater determinant,
while for the $^7\Delta_u$ and the $^7\Sigma_g^-$ states $n^*=N_\uparrow+1$, 
and  the role of non-dynamical correlations becomes crucial in our optimization of the AGP.
In fact the RVB provides in these latter cases  a remarkable energy gain 
of more than 1 eV as compared with a more simple but much less accurate JSD calculation, 
at least as far as the agreement with experiments is concerned.

\begin{table}
\begin{center}
\begin{threeparttable}

\caption{
\label{asymp_limits}
{\small  
Dissociation limits of various Fe$_2$ states for the RVB wave function.
$^5D$ (occupation $4s^2 3d^6$), and $^5F$ (occupation $4s^1 3d^7$) 
refer to neutral atom states, while $^4F$ (occupation $4s^2 3d^7$) 
is the ground state of the anion. The states denoted with $^{(2S+1)}[L_z]$
are non-definite angular momentum states for the presence of other components. 
They are eigenstate of $L_z$ but not of $L^2$.
In particular $^4$[0] and $^5$[0]  indicate the single occupation 
of the $d$ orbitals with $l_z=\pm 2$ and $l_z=0$,
$^5$[1] indicates the single occupation of the orbitals with $l_z=\pm 2$ and $l_z=-1$, while
the remaining $d$ orbitals are doubly occupied.
}
}
{ \small
\begin{tabular*}{\columnwidth}{@{\extracolsep{\fill}} l  
r @{+}@{\extracolsep{0pt}} l @{\extracolsep{\fill}} 
r @{+}@{\extracolsep{0pt}} l @{\extracolsep{\fill}} 
}
\hline
\hline
Fe$_2$ state & \multicolumn{2}{c}{from our wf} & \multicolumn{2}{c}{exact} \\ 
\hline
$^7\Sigma_g^- \rightarrow$ &  $^5D~$&$~^5$[1]  &   $^5D~$&$~^5F$ \\
$^9\Sigma_g^- \rightarrow$ &  $^5D~$&$~^5$[0]  &  $^5D~$&$~^5F$ \\
$^7\Delta_u  \rightarrow$  &  $^5D~$&$~^5$[0]  &  $^5D~$&$~^5D$ \\
\hline
$^8\Delta_g \rightarrow$    &  $^5D~$&$~^4$[0]  &  $^5D~$&$~^4F$ \\
$^8\Sigma_u^- \rightarrow$  &  $^5D~$&$~^4$[0]  &  $^5D~$&$~^4F$ \\
\hline
\hline
\end{tabular*}
}

\end{threeparttable}
\end{center}
\end{table}
All the lowest possible configurations  corresponding to  states 
obtained after the dissociation of our wave function are reported in Table~\ref{asymp_limits}. 
It is apparent that non-dynamical correlations are very important in the compound whenever
its total spin $S$ is less than the total maximum spin of the fragments.   
As anticipated, a controlled atomic limit can be obtained also in this case, 
with the caveat that  the total angular momentum $L_z$ around the 
molecular axis is conserved in the atomization process.
This implies that, within our RVB wave function, 
the total angular momentum $L$ of the fragments may not be definite in the atomization limit, 
and the corresponding JSD atomic 
reference energies depend explicitly on the angular momentum 
projection quantum number $L_z$.
In this way, 
we need to perform atomic calculations  
for various $L_z$ and choose the appropriate ones 
for the reference to the total energy of the dimer. 
The results of our atomic calculations are reported 
in Table~\ref{iron_atom_table}. The $^5D \rightarrow {^5F}$ transition is found to be very close to 
the experimental value, while the electron affinity is off by $0.3$ eV. This is largely due to
a lack of correlation energy in the LRDMC calculations for the anion, 
since MRCI calculations done with MOLPRO\cite{molpro} are in agreement with 
the experimental value. 
\begin{table}
\begin{center}
\begin{threeparttable}

\caption{
\label{iron_atom_table}
{\small  
The $^5F$, and $^5D$ energies
for the neutral atom, and the $^4F$ energy for the anion are reported in Hartree.
The non-definite angular momentum JSD states  are denoted  
with the conventions reported in Table~\ref{asymp_limits}.
They will be useful for the energy correction based on the dissociation limit. 
The variational wave function used here is a JSD like.
From the total energies of the spin-definite states 
we calculate the $^5D \rightarrow {^5F}$  excitations and 
the electron affinity, expressed in eV and compared with the experiment. 
}
}
{ \small
\begin{tabular*}{\columnwidth}{@{\extracolsep{\fill}} l  
r @{.}@{\extracolsep{0pt}} l @{\extracolsep{\fill}} 
r @{.}@{\extracolsep{0pt}} l @{\extracolsep{\fill}} 
}
\hline
\hline
 & \multicolumn{2}{c}{LRDMC} & \multicolumn{2}{c}{exp.} \\ 
\hline
$^5D$           &   -123&7819(11)  & \multicolumn{2}{c}{} \\ 
$^5F$           &   -123&7520(11)  & \multicolumn{2}{c}{} \\
$^5$[0]   &   -123&73986(72) & \multicolumn{2}{c}{} \\
$^5$[1]   &   -123&71856(75) & \multicolumn{2}{c}{} \\
$^4F$           &   -123&77731(94) & \multicolumn{2}{c}{} \\
$^4$[0]   &   -123&76544(81) & \multicolumn{2}{c}{} \\
\hline
$^5D \rightarrow {^5F}$ (eV) &  0&81(4) & 0&87 \tnote{a}   \\
$^4F \rightarrow {^5D}$ (eV) & -0&12(4) & 0&15 \tnote{a}  \\
\hline
\hline
\end{tabular*}
}

\begin{tablenotes}
{\footnotesize
\item[a]{From Ref.~\cite{moore49}}
}
\end{tablenotes}

\end{threeparttable}
\end{center}
\end{table}

After performing LRDMC simulations for the iron dimer at different interatomic distances (going from 3.5 to 8) 
and symmetry states, we found the results reported in Table~\ref{iron_dimer_table}.
The vibrational frequencies of $^9\Sigma_g^-$ and $^8\Sigma_g^-$ are in good
agreement with the Leopold's experimental data. 
Indeed, our best QMC estimate for the vibrational frequency
of the Fe$_2$ ground state is $\omega_e=301(15) cm^{-1}$, which 
matches perfectly the value 300(15) $cm^{-1}$ coming from PES,\cite{leopold86}
and the value 299.6 $cm^{-1}$ provided by Raman spectroscopy.\cite{moskovits80} 
Also for the anion dimer the LRDMC vibrational frequency ($\omega_e=210(20) cm^{-1}$) 
agrees with the experimental value of 250(25) $cm^{-1}$ yielded by PES.\cite{leopold86} 
Notice that our calculations correctly reproduce the
softening of the vibrational mode going from  $^9\Sigma_g^-$ to $^8\Sigma_g^-$.
Therefore, from these results we can confirm the symmetry of the peaks seen in
the PES. Indeed, the $^7\Delta_u$ state has
a much higher vibrational frequency, incompatible with the
experiment. It is interesting to highlight that the vibrational frequency for 
$^7\Delta_u$ computed with QMC simulations agrees with those
calculated by DFT methods for the same states.\cite{chretien02,gutsev03} 
\begin{table}[!ht]
\begin{center}
\begin{threeparttable}

\caption{
\label{iron_dimer_table}
{\small  
The LRDMC results are reported for some states of the neutral iron
dimer, and for the anion states $^8\Sigma_u^-$ and $^8\Delta_g$.   
We calculated total energies at the minimum of the interatomic
potential, equilibrium distances $R_e$, and
vibrational frequencies $\omega_e$. Some available experimental values
are also reported.
}
}
{ \small
\begin{tabular*}{\columnwidth}{@{\extracolsep{\fill}} l  
c c c c }
\hline
\hline
  & Energy & $R_e$ & $\omega_e$  & exp $\omega_e$  \\
  & (Hartree)   &  (a.u.)   &   ($cm^{-1}$) &  ($cm^{-1}$)        \\
\hline
$^7\Sigma_g^-$ &  -247.5036(20)   & 4.081(18)    &  327(15)  &  300(15) \tnote{b} \\
$^9\Sigma_g^-$ &  -247.5486(20)   & 4.093(19)    &  301(15)  &  299.6 \tnote{a} \\
$^7\Delta_u$   &  -247.5351(30)   & 3.894(18)    &  373(32)  &  -    \\ 
\hline
$^8\Delta_g$  &   -247.5585(30)   & 3.908(14)    &  354(24)  &  -    \\   
$^8\Sigma_u^-$ &  -247.5706(42)   & 4.276(28)    &  210(20)  & 250(20) \tnote{b} \\
\hline
\hline
\end{tabular*}
}

\begin{tablenotes}
{\footnotesize
\item[a]{From Ref.~\cite{moskovits80}}
\item[b]{From Ref.~\cite{leopold86}}
}
\end{tablenotes}

\end{threeparttable}
\end{center}
\end{table}
To check whether the $^9\Sigma_g^-$ is the true ground state, let us apply the correction 
to the energy levels based on the atomization limits.
To do so, we add to the total energy of the dimer the 
atomic energy differences between the right (the exact fully symmetric JSD limit) and the 
middle column (the asymptotic JSD limit 
that is possible to reach within our wave function) 
of Table~\ref{asymp_limits}.

The level ordering that we find is reported in Table~\ref{corrected_iron_dimer_table}.
It turns out that after the correction the $^7\Delta_u$ is the actual ground state, while the energy split 
between the $^9\Sigma_g^-$ and the $^7\Sigma_g^-$ states is $0.64(7)$ eV, in quite good agreement 
with the experimental findings ($0.534(4)$ eV).\cite{leopold86} 
The correction does not change the ordering for the anion, and so its ground state remains the $^8\Sigma_u^-$. 
This is a quite interesting effect of the correlation, which acts on the $d$ orbitals in a way which depends 
on the global symmetry and total charge of the system. This is apparent already at the atomic level, where the
occupation of the $d$ orbitals changes by going from the $^4F$ anion to the neutral $^5D$ ground states.
The energy difference between the anion $^8\Sigma_u^-$ state and
the neutral $^9\Sigma_g^-$ is $0.59(12)$ eV, which should be further shifted by $0.3$ eV due to the 
electron affinity correction in the atomic calculations, 
 because for this quantity the JSD is rather poor, as we have previously shown 
in Table~\ref{iron_atom_table}. This will lead to a difference of $\simeq 0.9$ eV
between the two states, again in agreement with the experimental value of $0.902(8)$ eV.\cite{leopold86}
This suggests that our approach can be further improved
by replacing the reference JSD atomization energies  
with exact atomic energies, readily available in quantum chemistry databases.
This approach is clearly useful and practical when one needs to consider total 
energy differences between electronic states 
with different particle number, like the electron affinity and ionization energies.
We would like to stress that, although the correction based on the asymptotic limits is ``approximate'', the accuracy level reachable in this way ($\approx 0.1$, maximum $0.2$ eV) 
is below the final energy differences.\cite{to_be_published}
Therefore, the ordering we propose here can be taken with a good confidence.

\begin{table}[!ht]
\begin{center}
\begin{threeparttable}

\caption{
\label{corrected_iron_dimer_table}
{\small  
The results for the energy minimum are reported for the same states 
as in Table~\ref{iron_dimer_table}, but with the correction 
described in the text which takes into account the atomic limit of the wave functions.
In the last two columns, we report the energy differences with respect to 
the state $^9\Sigma_g^-$ after the correction, and the experimental values taken 
from the Leopold's experiment.
}
}
{ \small
\begin{tabular*}{\columnwidth}{@{\extracolsep{\fill}} l  
c c c c}
\hline
\hline
 & Energy  & Corrected  & Difference  &  Exp \\
 & (Hartree)  &  (Hartree) & (eV) & (eV)  \\
\hline
$^7\Sigma_g^-$ &  -247.5036(20)   &  -247.5370(24)  &  +0.64(7)  & +0.534(4) \tnote{a} \\
$^9\Sigma_g^-$ &  -247.5486(20)   &  -247.5608(24)  &   0.0   &   \\  
$^7\Delta_u$   &  -247.5351(30)   &  -247.5771(33)  &  -0.44(9) & -  \\
\hline
$^8\Delta_g$  &   -247.5585(30)   &  -247.5703(33)  &  -0.26(9) & - \\   
$^8\Sigma_u^-$ &  -247.5706(42)   &  -247.5824(44)  &  -0.59(12) & -0.902(8) \tnote{a} \\
\hline
\hline
\end{tabular*}
}

\begin{tablenotes}
{\footnotesize
\item[a]{From Ref.~\cite{leopold86}}
}
\end{tablenotes}

\end{threeparttable}
\end{center}
\end{table}
Notice that the determination of the $^7\Delta_u$ as the ground state is also supported by
the equilibrium bond lengths provided by our calculations. Indeed, the experimental 
bond length is $3.82(4)$,\cite{purdum} which is very close to our findings for the $^7\Delta_u$ 
(see Table~\ref{iron_dimer_table}). On the other hand, 
the bond length of the anion has been measured only indirectly, since the unique
available data are taken from the PES, which revealed a
\emph{variation} of the equilibrium distance during the excitation from the
anion to the neutral iron dimer. A harmonic Franck-Condon analysis of
the vibronic band intensity profile yielded a bond elongation of 0.15(4)
a.u. on electron attachment.\cite{leopold86} Now, the difference between the
$^9\Sigma_g^-$ and $^8\Sigma_u^-$ bond length of our LRDMC calculation amounts to 0.18 a.u.,
which is in perfect agreement with the elongation of the anion dimer measured
in the PES.
Therefore, we conclude that the states seen in the Leopold's photoelectron spectrum are 
the anion $^8\Sigma_u^-$, and the neutral $^9\Sigma_g^-$ and $^7\Sigma_g^-$. The ground state
of the neutral dimer is however the $^7\Delta_u$ state, not seen in the photoelectron experiment
since the transition from a $^8\Sigma_u^-$ 
(the anion symmetry of the prepared initial state) to a $^7\Delta_u$ symmetry 
is a second-order process, and so of negligible rate with respect to 
the $^9\Sigma_g^-$ and $^7\Sigma_g^-$ states, connected with the $^8\Sigma_u^-$
by a direct photodetachment of the electron living in the $4 \sigma^* (4s)$ orbital.\cite{leopold88}

\section{Conclusions}

We have shown that
by using a cheap and nevertheless very accurate realization 
of the RVB wave function based on the MO expansion 
of the AGP part, it is possible to tackle
highly debated and challenging  
transition metal compounds.
The solution of the iron dimer puzzle appears at 
end, and we strongly believe that many other problems -
where the electron correlaton plays a strong role -
could be finally understood within this framework.

\section*{Acknowledgments}

This work was partially supported by COFIN2007, and CNR. 
One of us (M.C.) acknowledges support in the form of 
the NSF grant DMR-0404853 during his stay at the University of Illinois 
at Urbana-Champaign, and thanks the Centre de Physique Th\'eorique of the
Ecole Polytechnique, where this work was partially accomplished.
We acknowledge useful discussions with N. Marzari, H. Kulik,  L. Mitas, and L. Guidoni.

\bibliographystyle{elsarticle-num}
\bibliography{elscpl}

\end{document}